\documentclass[aps,prd,floatfix,superscriptaddress,twocolumn,showpacs]{revtex4}

\usepackage{epsfig}
\usepackage{graphicx,psfrag}
\usepackage{dcolumn}
\usepackage{bm}

\newcommand{\beq}{\begin{equation}}
\newcommand{\eeq}{\end{equation}}
\newcommand{\bea}{\begin{eqnarray}}
\newcommand{\eea}{\end{eqnarray}}
\newcommand{\hf} {\frac{1}{2}}
\newcommand{\nonu}{\nonumber\\}
\newcommand{\nn}{\nonumber\\}

\newcommand\eqn[1]     {Eq.\,(\ref{#1})}

\newcommand\fig[1]     {Fig.\,{\ref{#1}}}

\newcommand\tab[1]     {Table~\ref{#1}}

\def\eq#1{(\ref{#1})}
\def\s0#1#2{\mbox{\small{$ \frac{#1}{#2} $}}}
\def\0#1#2{\frac{#1}{#2}}

\def\tu{{\tilde u}}

\def\ord#1{{\cal O}(#1)}

\def\mr#1{{\mathrm{#1}}}

\begin{document}

\title{Bosonization and Functional Renormalization Group Approach\\ 
in the Framework of QED$_2$}

\author{I. N\'andori}
\affiliation{Institute of Nuclear Research, P.O.Box 51, 
H-4001 Debrecen, Hungary} 

\begin{abstract} 
Known results on two-dimensional quantum electrodynamics (QED$_2$) 
have been used to study the dependence of functional renormalization group 
equations on renormalization schemes and approximations applied for its 
bosonized version. It is demonstrated that the singularity of flow equations 
can be avoided in the optimized and power-law schemes for the bosonized 
model and the drawback of renormalization on bosonization is shown: 
it is indicated that renormalization of QED$_2$ possibly requires interaction 
terms corresponding to higher frequency modes of its bosonized version.
\end{abstract}

\pacs{11.10.Hi, 11.10.Gh, 11.10.Kk}

\maketitle

\section{Introduction}
\label{sec_intro}
In low dimensions, bosonization rules enable one to reformulate fermionic and 
gauge models in terms of elementary scalar fields. For example, the bose form 
of the two-dimensional massive Thirring model \cite{thirring} is the massless 
sine-Gordon (SG) scalar theory \cite{sg_coleman}. Single flavor QED$_2$ with 
massive fermions can also be bosonized and the corresponding scalar theory is 
the massive sine-Gordon (MSG) model \cite{msg_coleman,general_msg,msg}. 
Furthermore, the multiflavor QED$_2$, and the two-dimensional multicolor 
quantum chromodynamics (QCD$_2$) can also be rewritten as multicomponent 
SG theories \cite{,general_msg,msg,qed_qcd}. Thus, corresponding bose models 
are usually SG-type theories. The critical behavior of original fermionic and gauge 
theories and their bosonized versions can be studied by various methods such as 
renormalization group (RG) approaches. If critical behaviors of fermionic and gauge 
models are known, then bosonization transformations can be used to consider the 
dependence of methods suitable for the critical behavior of SG-type models on the 
approximations used.

The goal of this paper is to study the dependence of functional RG equations 
obtained for the MSG model on renormalization schemes and applied 
approximations. The phase structure of the corresponding fermionic theory, 
the single flavor QED$_2$ with massive fermions has already been mapped 
out by density matrix RG approach \cite{dmrg_critical} which is used here to 
optimize the scheme-dependence for the MSG model. The drawback of the RG 
study of the MSG model on the bosonization transformations is also discussed, 
namely it is indicated that renormalization of QED$_2$ possibly requires 
interaction terms corresponding to higher frequency modes of the MSG model.

The paper is organized as follows. In Sec.~\ref{sec_bose}, some aspects of 
bosonization are discussed. Functional RG equations are obtained for the MSG 
model in the second order of the gradient expansion in Sec.~\ref{sec_func}. 
Results of perturbative RG are summarized briefly in Sec.~\ref{sec_perturb}. 
The nonperturbative RG study is given in Sec.~\ref{sec_single} for the 
single-frequency, and in Sec.~\ref{sec_multi} for the multi-frequency MSG 
model in local potential approximation (LPA). In Sec.~\ref{sec_beyond_lpa}, 
the RG flow is determined beyond LPA. Section~\ref{sec_sum} serves as the
summary.

\section{Bosonization}
\label{sec_bose}
The mapping of quantum field theories of interacting fermions onto 
an equivalent theory of interacting bosons called bosonization is 
well-established in the context of $1+1$-dimensional theories. The 
well-known example is the massive Thirring model \cite{thirring}, which 
is a theory of a single Dirac field $\psi$ determined by the Lagrangian 
density
\begin{equation}  
\label{thirring}
{\cal L}_{\mathrm{Thirring}} =  
{\bar\psi} (i \gamma^{\mu} \partial_{\mu} - m) \psi  
- \hf g j^{\mu} j_{\mu}   
\end{equation}
where $j^{\mu} = {\bar\psi} \gamma^{\mu} \psi$, $m$ is the mass, and 
$g$ is the coupling. This can be mapped onto the SG scalar field theory 
described by the Lagrangian density \cite{sg_coleman}
\begin{equation}  
\label{sg}
{\cal L}_{\mathrm{SG}} =  
\hf \, (\partial_{\mu} \varphi)^2 
+ u \cos(\beta \, \varphi)  
\end{equation}
where $\varphi$ is a one-component scalar field 
and the identifications $4\pi/\beta^2  = 1+ g/\pi$, 
$- \beta/(2\pi) \epsilon^{\mu\nu} \partial_{\nu} \varphi  =  j^{\mu}$ and
$u \cos(\beta\varphi)  = -m \bar\psi \psi$, are made between the parameters 
of the two models. (Conventions and the definition for appropriate normal 
ordering are given in \cite{sg_coleman}.) The Lagrangian of QED$_2$ with 
a massive Dirac fermion which is also  called the massive Schwinger model 
reads as \cite{msg_coleman}
\begin{equation}
\label{qed_2}
{\cal{L}}_{\mathrm{QED_2}} = {\bar\psi}
\left(i \gamma^{\mu} \partial_{\mu} - m - e\gamma^{\mu}  A_{\mu} \right)
\psi -\frac{1}{4} F_{\mu\nu} F^{\mu\nu}
\end{equation}
where $F_{\mu\nu} = \partial_{\mu}A_{\nu} - \partial_{\nu} A_{\mu}$. Using 
bosonization technique the fermionic theory \eq{qed_2} can be mapped 
onto an equivalent Bose form \cite{msg_coleman} which is considered as 
the specific form of the MSG model \cite{scheme_sg,general_msg,msg}
whose Lagrangian density is written as 
\begin{equation}
\label{msg}
{\cal{L}}_{\mathrm{MSG}} = \hf (\partial_{\mu} \varphi)^2
+ \hf M^2 \varphi^2
+ u \cos (\beta \varphi)
\end{equation}
with $\beta^2= 4\pi$, $M^2 = e^2/\pi$, $u = e \, m\, \exp{(\gamma})/(2\pi^{3/2})$
where $\gamma = 0.5774$ is the Euler's constant and the vacuum angle 
parameter has to be chosen as $\theta =\pm \pi$ for $u>0$ and $\theta =0$ 
for $u<0$ \cite{general_msg}. The MSG model has two phases. The Ising-type 
phase transition \cite{dmrg_critical} is controlled by the dimensionless quantity 
$u/M^2$ related to the critical ratio $(m/e)_c$ of QED$_2$ which separates the 
confining and the half-asymptotic phases of the fermionic model. The critical ratio 
$(m/e)_c = 0.31-0.33$ has been calculated by the density matrix RG method for 
the fermionic model which implies \cite{dmrg_critical}
\bea
\label{exact_ratio}
\left(\frac{u}{M^2}\right)_c  
=   \left(\frac{m}{e}\right)_c \frac{\exp{(\gamma)}}{2\sqrt{\pi}}
=   0.156 - 0.168.
\eea
If one assumes a quartic self-interaction among the massive Dirac fermions 
of QED$_2$ by adding a Thirring type term to the Lagrangian \eq{qed_2} 
then one arrives at the massive Schwinger-Thirring model which reads as
\begin{equation}
\label{mst}
{\cal{L}} = {\bar\psi}
\left(i \gamma^{\mu} \partial_{\mu} - m - e\gamma^{\mu}  A_{\mu} \right)
\psi -\frac{1}{4} F_{\mu\nu} F^{\mu\nu} -\hf g j^{\mu} j_{\mu}
\end{equation}
where $F_{\mu\nu} = \partial_{\mu}A_{\nu} - \partial_{\nu} A_{\mu}$ and 
$j^{\mu} = {\bar\psi} \gamma^{\mu} \psi$. It has been argued that by using 
bosonization technique the fermionic theory \eq{mst} can be mapped onto 
the MSG model if  $4\pi/\beta^2 = 1 + g/(2\pi)$ and the Fourier amplitude is 
related to the fermion mass ($u\sim m$) and $M^2 = e^2/(\pi + g/2)$. Let us
note that the bosonization of two-dimensional gauge and fermionic models
(special attention on the Schwinger-Thirring model) has been the subject of 
intense study \cite{bosonization_msg}.

\section{Functional RG method for the MSG model}
\label{sec_func}
In this section, we derive functional RG equations for the MSG model.
Recently, the complete phase structure of the SG model \eq{sg} has been 
mapped out by extending the functional RG analysis \cite{rg_sg} beyond 
LPA \cite{sg_prl}. Here we use the same RG approach for the MSG model 
\eq{msg}. Namely, the effective average action functional RG method 
\cite{We1993,Mo1994,internal} where the evolution equation reads as 
\beq
\label{feveq}
k\partial_k\Gamma_k=\hf\mr{Tr}  [(R_k+\Gamma''_k)^{-1}  \, k\partial_k R_k]
\eeq
with the notation $^\prime=\partial/\partial\varphi$, and the trace Tr stands 
for the integration over all momenta. As exact RG Eqs. \eq{feveq} are 
functional equations they are handled by truncations. Truncated RG flows 
depend on the choice of the regulator function $R_k$, i.e. on the 
renormalization scheme. In order to optimize the scheme-dependence 
various strategies have been worked out. For example, a general optimization
procedure was proposed to increase the convergence of the truncated flow
\cite{opt_rg}, and successfully applied in many cases, e.g. in quantum gravity 
\cite{opt_qg} or in low-energy QCD \cite{opt_qcd}. Although, the optimized 
regulator $R^{\mr{opt}}_k$ \cite{opt_rg} does not support the gradient expansion 
beyond second order \cite{Ro2010, Mo2005} but an optimization criterion based 
on functional variation is proposed to handle this problem \cite{opt_func} and 
has been used for the study of low-energy behavior of the Yang-Mills theory. Since 
the RG study of the SG model was done by using the power-law type regulator 
$R^{\mr{pow}}_k$ \cite{Mo1994}, it is a natural choice that for the RG analysis
of the MSG model we use also the power-law regulator and the previously 
mentioned optimized one
\beq
\label{regulator}
R^{\mr{pow}}_k = p^2\left(\frac{k^2}{p^2}\right)^b, \,\,\,
R^{\mr{opt}}_k = a (k^2 -p^2) \Theta(k^2 -p^2)
\eeq
where $b\ge 1$, usually $a=1$ and $\Theta(x)$ is the Heaviside step-function. 
Let us note that various types of regulator functions can be used (e.g. the
exponential one \cite{We1993}) but here we focus on those \eq{regulator} 
which provide us the possibility to perform momentum integrals of RG equations 
analytically.  

Another problem related to truncations is the singularity of RG flows. The 
appearance of spinodal instability (SI), i.e. singularity in the RG flow could be 
the consequence of a too drastic truncation. In this work we show that the 
singularity of the RG flow obtained for the MSG model can be avoided for the 
optimized and power-law regulators \eq{regulator}, if the RG equation obtained 
in LPA is integrated out directly i.e. using appropriate approximations the effective 
potential remains convex \cite{convexity}. Let us note that the MSG model has 
already been investigated by using the truncated Fourier expansion of the 
potential which was found to be a too drastic simplifications of the functional 
subspace since SI appeared in the RG flow and the Maxwell construction 
resulted in a degenerate effective potential scheme-independently 
\cite{scheme_sg}. 

Equation \eq{feveq} has been solved over the functional subspace spanned 
by the ansatz for the MSG model \eq{msg}
\beq
\label{eaans}
\Gamma_k = \int_x\left[\frac{1}{2} z (\partial_\mu\varphi_x)^2+V_k(\varphi_x)\right],
\eeq
where the local potential contains a single Fourier mode 
\beq
\label{single_mode}
V_k(\varphi) =  
\hf {\bf M}^2(k) \,\, \varphi^2 + u(k) \cos(\varphi),
\eeq
and the following notations are introduced
\beq
\label{identifications}
{\bf  M}^2 \equiv z \, M^2,  \hskip 0.5cm
z \equiv 1/\beta^2
\eeq
via the rescaling of the field $\varphi \to \varphi/\beta$ in \eq{msg} and 
$z(k)$  stands for the field-independent wave-function renormalization.
Although RG transformations generate higher harmonics, we use the 
simple ansatz  \eq{single_mode} first since in the case of the SG model 
it was found to be an appropriate approximation \cite{sg_prl}. Then 
\eqn{feveq} leads to the evolution equations \cite{sg_prl}
\bea
\label{ea_v}
\partial_k V_k &=& \hf\int_p{\cal D}_kk\partial_k R_k,\\
\label{ea_z}
k\partial_kz &=&
{\cal P}_0 V'''^2_k\int_p{\cal D}_k^2k\partial_k R_k\left(
\frac{\partial^2{\cal D}_k}{\partial p^2\partial p^2}p^2
+\frac{\partial{\cal D}_k}{\partial p^2}
\right)
\eea
with ${\cal D}_k=1/(zp^2+R_k+V''_k)$ and 
${\cal P}_0=(2\pi)^{-1}\int_0^{2\pi} d\varphi$ as the projection onto the 
field-independent subspace. The scale $k$ covers the momentum 
interval from the UV cutoff $\Lambda$ to zero. Inserting the ansatz 
\eq{single_mode} into Eqs. \eq{ea_v} and \eq{ea_z} the flow equations 
for the coupling constants are (similar RG equations obtained for the SG 
model in \cite{rg_sg}) 
\bea
\label{general_ea_u}
k\partial_k u &=&
\frac1{2\pi} \int_p ~\frac{p (k\partial_k R_k)}{u}\left(\frac{P}{\sqrt{P^2-u^2}}-1\right),\\
\label{general_ea_z}
k\partial_k z &=& \frac1{2\pi}\int_p p (k\partial_k R_k)
\biggl(\frac{u^2 p^2 (\partial_{p^2}P)^2(4P^2+u^2)}{4(P^2-u^2)^{7/2}}\nn
&&-\frac{u^2P(\partial_{p^2}P+p^2\partial_{p^2}^2P)}{2(P^2-u^2)^{5/2}} \biggr)
\eea
with $P = zp^2+{\bf M}^2 +R_k$ where $\partial_k {\bf M} =0$. In general, 
the momentum integrals have to be performed numerically, however, in some 
cases analytical results are available. Indeed, by using the power-law type
regulator function with $b=1$ (i.e. the Callan-Symanzik scheme), the momentum 
integrals can be performed and the RG equations reads as, 
\bea
\label{single_b1_exact}
(2+k\partial_k)\tu &=& 
\frac{1}{2\pi z \tu}  
\left[1+{\tilde{\bf M}}^2 -  \sqrt{(1+{\tilde{\bf M}}^2)^2 - \tu^2} \right] \nn
k\partial_kz &=& 
-\frac{1}{24\pi} \frac{\tu^2}{[(1+{\tilde{\bf M}}^2)^2 - \tu^2]^\frac{3}{2}} \nn
(2+k\partial_k)  {\tilde{\bf M}}^2 &=& 0,
\eea
with dimensionless couplings $\tu = k^{-2} u$,
${\tilde{\bf M}}^2 = k^{-2} {\bf M}^2$.

\section{Perturbative RG}
\label{sec_perturb}
Let us first consider the massless limit ($\tilde{\bf M} \to 0$) when the RG 
Eqs.~\eq{single_b1_exact} reduce to those derived for the SG model 
\cite{sg_prl}. The spontaneously broken phase of the SG model is known to be 
equivalent to the neutral sector of the massive Thirring model \eq{thirring}. 
Indeed, the renormalization of the massive Thirring model has already been 
discussed and it was demonstrated that the scaling of the fermion mass (see, 
e.g. \cite{zj}), $m = m_0 (\mu/\Lambda)^{(-g/(g+\pi))}$ is identical to the solution 
of the linearized form of the RG Eqs.~\eq{single_b1_exact} if $k\partial_k z =0$ 
and $\tilde{\bf M} =0$ are assumed and the equivalences $g/(g+\pi) = 1-1/(z 4\pi)$, 
$k=\mu$, $u = m \mu/\pi$ and $u_\Lambda = m_0 \Lambda/\pi$ are used.

For nonvanishing mass ($\tilde{\bf M} \neq 0$), it is illustrative to compare 
the perturbative RG equations given in \cite{IcMu1994,PiVa2000} to that  
obtained by the linearization of  Eq. \eq{single_b1_exact} which reads as
\bea
\label{single_b1_lin}
(2+k\partial_k)\tu &=&  
 \frac{1}{4\pi}  \frac{\tu}{z}  \frac{1}{1+{\tilde{\bf M}}^2}  + \ord{\tu^2} \nn
k\partial_kz &=&   
-\frac{1}{24\pi} \tu^2 \frac{1}{(1+{\tilde{\bf M}}^2)^3}  + \ord{\tu^3} \nn
(2+k\partial_k)  {\tilde{\bf M}}^2 &=& 0,
\eea
Let us relate the parameters of  \cite{IcMu1994} 
$\mu$, $\kappa$ and $m^2$ to that of defined in Eqs. \eq{eaans} and 
\eq{single_mode}, namely $u = \mu$, $z = 1/(4\pi^2 \kappa)$, 
${\bf M}^2 = m^2/(4\pi \kappa)$ and $k = \Lambda$. By using these 
equivalences, the approximated RG equations (3.17) of \cite{IcMu1994} 
can be written as 
\bea
\label{ichinose}
(2+k\partial_k)\tu &=&   
\frac{1}{4\pi}  \frac{\tu}{z}  \frac{z}{z+{\tilde{\bf M}}^2}  
+ \frac{1}{16\pi}  \frac{\tu^3}{z^3} \left( \frac{z}{z+{\tilde{\bf M}}^2}\right)^3  
 \nn
k\partial_kz &=& 
-\frac{\alpha_2}{8}    \frac{\tu^2}{z}  \frac{z}{z+{\tilde{\bf M}}^2}  \nn
(2+k\partial_k)  {\tilde{\bf M}}^2 &=& 0,
\eea
with $\alpha_2$ constant. Perturbative results (Eqs. \eq{ichinose} and 
\eq{single_b1_lin}) demonstrate that the 
Kosterlitz-Thouless-Berezinski \cite{KTB} (KTB) type phase transition known 
to take place for the SG theory \cite{rg_sg}, disappears in case of the MSG 
model.  Indeed, one expects an Ising-type second order phase transition 
for the MSG scalar model which is controlled by the dimensionless ratio 
$u/M^2 = \tu/{\tilde M^2}$. For example, according to Eq. \eq{single_b1_lin}, 
in the infrared (IR) limit ($k\to 0$) the mass term becomes large 
$\tilde{\bf M}^2 \sim k^{-2}$ which freezes out the evolution of $z$ and the 
Fourier amplitude has a trivial tree-level scaling $\tu \sim k^{-2}$; consequently, 
the ratio $\tu/{\tilde M}^2$ tends to a constant value.
 
However, the perturbative RG flow produces arbitrary large ratio depending 
on the initial conditions (there is no upper bound), hence no critical value 
can be determined. Consequently, the perturbative RG flow is not suitable 
for the prediction of the Ising-type phase transition of the MSG  model.

\section{Functional RG study of the single-frequency MSG model}
\label{sec_single}
\begin{figure}[ht] 
\begin{center} 
\epsfig{file=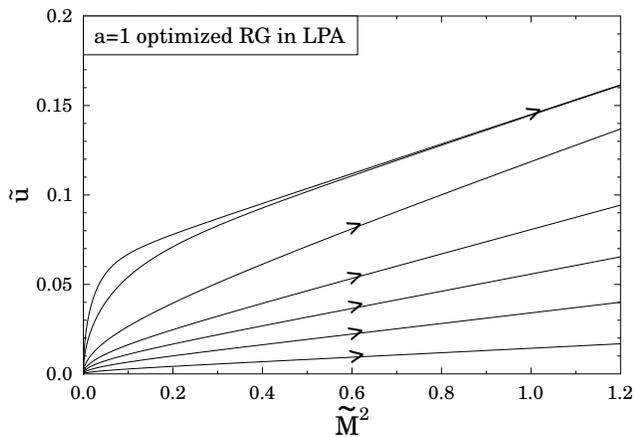,width=8.3 cm}
\caption{
\label{opt_phase}
Phase diagram of the MSG model for $\beta^2 =4\pi$. RG trajectories 
are obtained by the integration of Eq. \eq{optimized} . Since SI does
not occur in the RG flow, the critical ratio of the MSG model can be
determined, $[{\tilde u}/{\tilde M^2}]_c = 0.0625$. The arrows indicate 
the direction of the flow.
} 
\end{center}
\end{figure}

\begin{figure}[ht] 
\begin{center} 
\epsfig{file=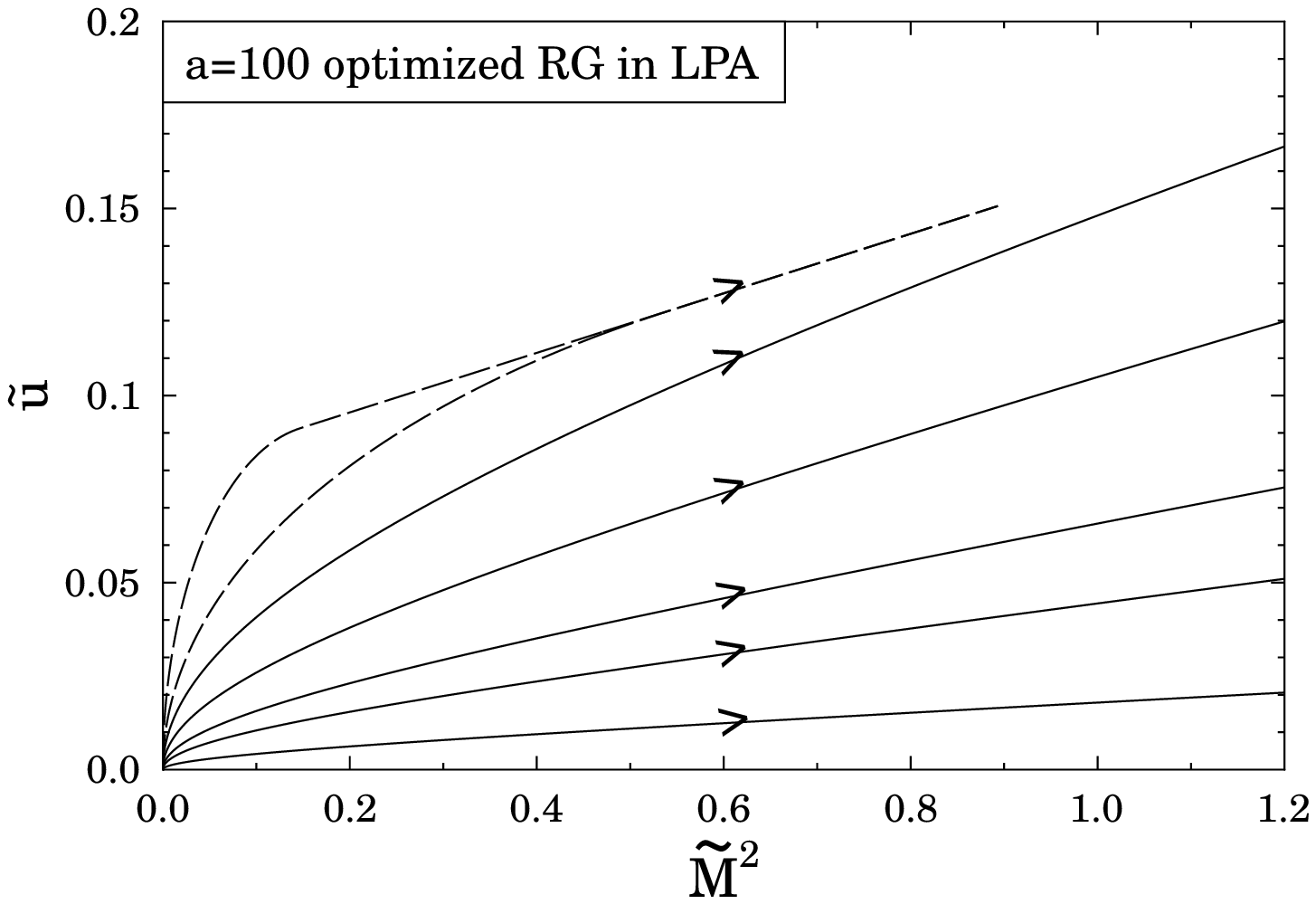,width=8.3 cm}
\caption{
\label{opt_a100}
Phase diagram of the MSG model for $\beta^2 =4\pi$. RG trajectories are 
obtained by the integration of Eq. \eq{opt_a} with $a=100$. The dashed 
lines correspond to RG trajectories where SI occurs in the RG flow, thus 
the critical ratio of the MSG model cannot be obtained. 
} 
\end{center}
\end{figure}

\begin{figure}[ht] 
\begin{center} 
\epsfig{file=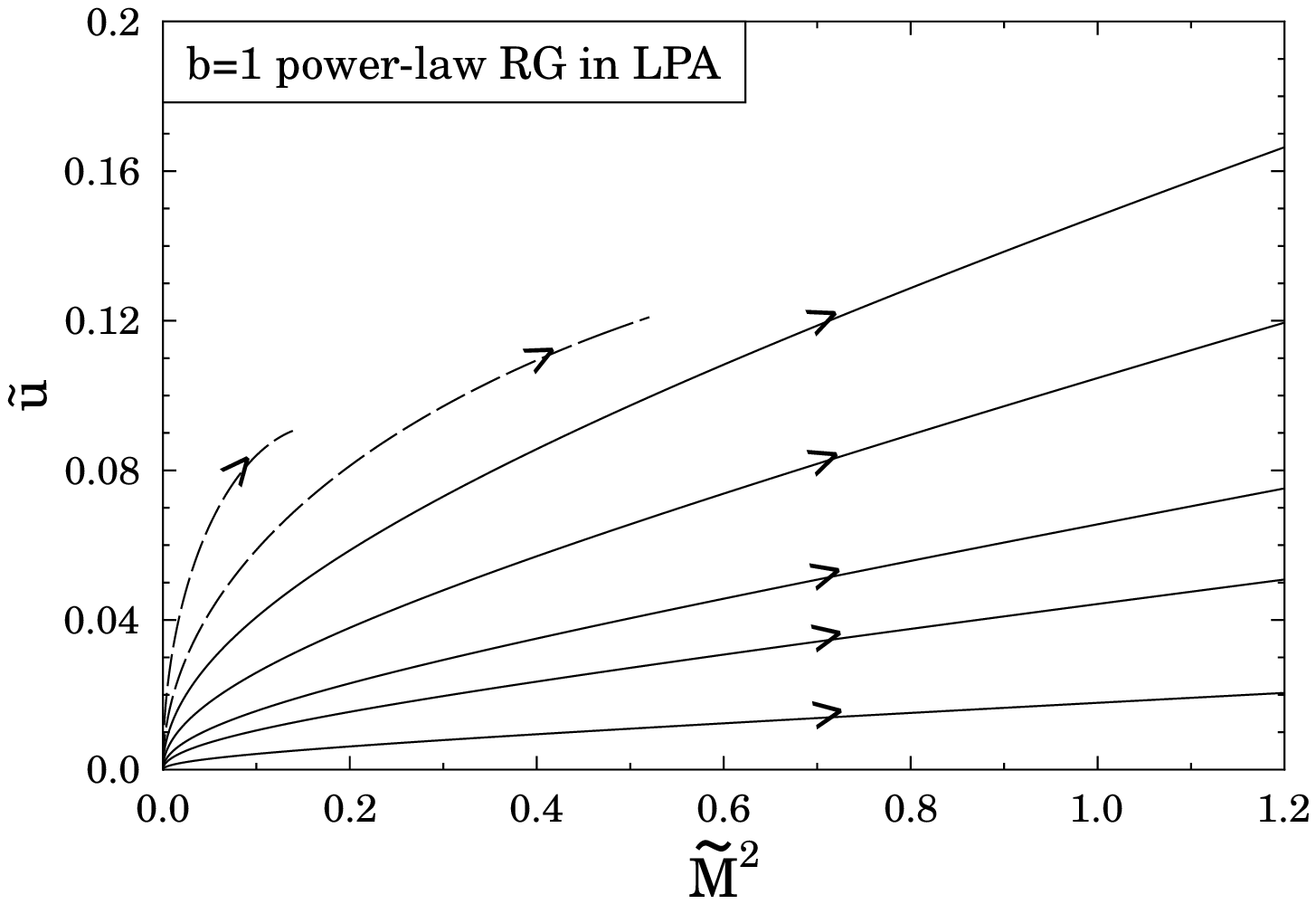,width=8.3 cm}
\caption{
\label{b1_lpa_phases}
Phase diagram of the MSG model for $\beta^2 =4\pi$. RG trajectories are 
obtained by the integration of Eq. \eq{b1_lpa}. The dashed lines correspond 
to RG trajectories where SI occurs in the RG flow, thus the critical ratio of the 
single-frequency MSG model cannot be obtained. 
} 
\end{center}
\end{figure}

Exact RG equations (with a single Fourier mode) show a different picture 
since one cannot obtain arbitrary large IR value for the ratio $\tu/{\tilde M}^2$. 
Let us first consider the RG flow obtained by the optimized regulator function 
\eq{regulator} with $a=1$ in the LPA (i.e. $z=1/\beta^2=$constant) which reads 
as
\bea
\label{optimized}
(2+ k \partial_k) \tu &=& - \frac{1}{2\pi\beta^2 \tu}
\left[1-\sqrt{\frac{(1+{\tilde{M}}^2)^2}{(1+{\tilde{M}}^2)^2 + \beta^4 \tu^2}}\right],
\nonu
(2+k\partial_k)  {\tilde{M}}^2 &=& 0.
\eea
In the broken symmetric phase, the RG trajectories merge into a single 
trajectory in the deep IR region which is characterized by the critical ratio 
$[\tu/\tilde M^2]_c =  0.0625$ (for $\beta^2 =4\pi$) and serves as an upper 
bound (see \fig{opt_phase}). 
%
%
The critical value obtained by Eq. \eq{optimized} is less than the exact 
result \eq{exact_ratio}, therefore it requires further improvement. Let us try
to improve it by the optimized regulator $\eq{regulator}$ with $a\neq 1$ 
which has the following form in LPA
\bea
\label{opt_a}
&&(2+ k \partial_k) \tu = \frac{a}{(a-1)2\pi\tu \beta^2}
\left[(1+ {\tilde M^2}) - (a + {\tilde M^2})  \right. \nonu
&&\left. + \sqrt{(a+ {\tilde M^2})^2 -\tu^2 \beta^4} 
- \sqrt{(1+ {\tilde M^2})^2 -\tu^2 \beta^4} \right],
\nonu
&&(2+k\partial_k)  {\tilde{M}}^2 = 0.
\eea
However, for $a\neq 1$ spinodal instability (SI) appears in the RG flow in 
the broken symmetric phase, i.e. RG equations become singular in the IR 
limit and the RG flow stops at some finite scale (see the dashed lines in 
\fig{opt_a100}). Although RG trajectories start to converge into a single one 
in the broken phase the critical value of the single-frequency MSG model 
cannot be determined unambiguously. In other words, the convergence 
properties of the optimized RG is weakened for $a\neq 1$.  
%
%
Let us try to use other types of RG equations, for example the power-law
type RG with $b=1$ in LPA which reads as
\bea
\label{b1_lpa}
&&(2+ k \partial_k) \tu = 
\frac{(1+ {\tilde M^2}) - \sqrt{(1+ {\tilde M^2})^2 -\tu^2 \beta^4} }{2\pi \tu \beta^2},
\nonu
&&(2+k\partial_k)  {\tilde{M}}^2 = 0.
\eea
The RG Eq.~\eq{b1_lpa} can be obtained by rescaling $\tu \to \tu/z$ 
and  ${\tilde {\bf M}}^2 \to {\tilde {\bf M}}^2/z$ in \eq{single_b1_exact} and 
using the identifications \eq{identifications} with the assumption $\partial_k z=0$. 
It is known that in the sharp limit, the optimized RG becomes identical to the 
power-law type RG with $b=1$, i.e. Eq. \eq{opt_a} reduces to Eq. \eq{b1_lpa} 
for $a\to \infty$. Thus, SI is expected in case of Eq. \eq{b1_lpa}.  Indeed, the 
numerical solution of \eq{b1_lpa} indicates the appearance of SI in the broken 
symmetric phase (see the dashed lines in \fig{b1_lpa_phases}). 
%
%

\begin{figure}[ht] 
\begin{center} 
\epsfig{file=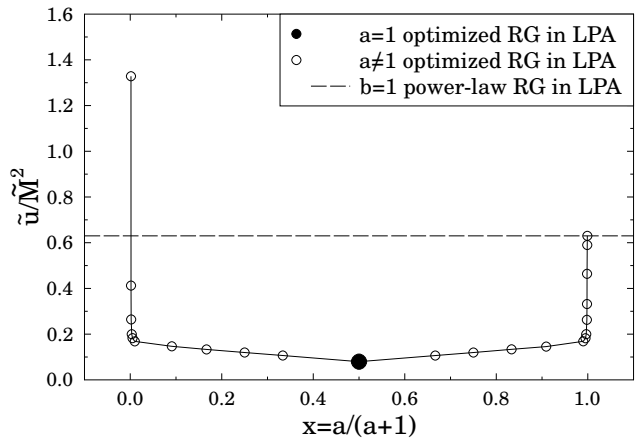,width=8.3 cm}
\caption{
\label{compare}
This figure shows how the IR value of the ratio $\tilde u/\tilde M^2$ 
obtained by the integration of RG Eqs.~\eq{optimized} and 
\eq{opt_a} depends on the parameter $a$ of the regulator function. 
The same initial condition has been used for the numerical integration, 
$({\tilde u}(\Lambda)= 10^{-5}, {\tilde M^2}(\Lambda)=10^{-9})$. SI has 
occured for $a\neq 1$, thus the RG flow stops at some finite scale 
where the ratio has been read off and plotted. It is possible to avoid SI
but only for $a=1$. For $a\to \infty$ (i.e. $x\to 1$) the RG Eq.~\eq{opt_a} 
becomes identical to that was obtained by the power-law type 
regulator with $b=1$ \eq{b1_lpa}; consequently in this case the IR values 
of the ratio coincide.
} 
\end{center}
\end{figure}

It is also illustrative to compare the IR values of the ratio $\tu/\tilde M^2$ at the 
scale of SI given by the integration of optimized RG with various values for the 
parameter $a$ using the same UV initial condition (see \fig{compare}). This 
demonstrates that the best estimate for the critical ratio of the single-frequency 
MSG model, in the framework of the optimized RG can be achieved for $a=1$. Our 
findings are consistent to the feature of the optimized RG namely that it increases 
the convergence properties of the truncated flow. For example, similar result is 
shown in Fig.~12 of \cite{litim_o(n)} in the framework of the O(N) symmetric scalar 
theory in $d=3$ dimensions which has been the subject of intense study on 
scheme-dependence (see e.g. \cite{scheme}). Let us note that the optimized 
RG with $a=1$ produces reliable results for the single-frequency MSG model in 
LPA as opposed to the Wilson-Polchinski RG \cite{polch} which was found to
be inappropriate \cite{scheme_sg} for the determination of the phase structure 
of the MSG model. This is a counterexample for the statement, namely, that the 
Wilson-Polchinski RG and the optimized one always provide us with the same 
critical behavior in LPA \cite{Mo2005}. The mapping between the two latter RG 
methods works only if the potential is nondegenerate but this is not true for the 
broken symmetric phase of the MSG model.
%
%

Since Eq.~\eq{optimized} has no singular behavior, the appearance of SI 
is expected to be the consequence of an inappropriate approximation, e.g. 
too drastic simplification of the functional subspace. Let us also note that 
SG-type models undergo an infinite order (or topological) phase transition 
and it was shown \cite{sg_prl} that the study of a single-frequency model 
is sufficient to recover the KTB-type critical properties (higher harmonics were 
found to be irrelevant). However, the MSG model has an Ising-type second
order phase transition, hence there is no reason to focus on the study of
a single-frequency model. Consequently, in order to obtain reliable results 
and to avoid SI in case of the MSG model one has to incorporate higher 
harmonics generated by RG Eqs.~\eq{ea_v}, \eq{ea_z} which is 
discussed in the next section.

\section{Functional RG study of the multi-frequency MSG model}
\label{sec_multi}
We now turn to the discussion of the MSG model including higher 
harmonics. Let us consider the RG Eq.~\eq{ea_v} in LPA
(i.e. $\partial_k z =0$) using the ansatz 
\beq
\label{multi}
V_k(\varphi) =  
\hf M^2  \varphi^2 + \sum_{n=1}^{\infty} u_{n}(k) \cos(n\beta \varphi).
\eeq
There are two ways to determine the IR scaling of the MSG model (i) either 
the partial differential Eq.~\eq{ea_v} has to be solved directly by e.g. a 
computer algebraic program using the initial condition \eq{msg} (higher 
harmonics are generated by RG equations), (ii) or one can find the solution of 
ordinary differential equations given for the coupling constants which are 
obtained by inserting the ansatz \eq{multi} into Eq.\eq{ea_v}.
Let us first discuss the latter case when it is unavoidable to implement a 
further approximation besides the LPA, namely the 
truncation of the Fourier expansion of the potential. In this case, the RG flow 
on the trajectories started at $\beta^2=4\pi$ in the broken symmetric phase
develops SI at some finite scale $k_{\mr{SI}}>M$ \cite{scheme_sg}. The way 
to go beyond $k_{\mr{SI}}$ is the Maxwell construction (i.e. the tree-level 
blocking relation \cite{tree}) which represents too strong constraint on the RG 
flow and results in a scheme-independent infrared value for the critical ratio 
$[\tu/\tilde M^2]_c = 0.159$ \cite{scheme_sg}. However, SI  occurs in the 
RG flow as an artifact \cite{scheme_sg,direct_rg} due to the truncated 
Fourier-expansion applied to the almost degenerate blocked action of the 
MSG model, thus, one has to solve directly the RG Eq.~\eq{ea_v} in 
order to obtain reliable results.

\begin{figure}[ht] 
\begin{center} 
\epsfig{file=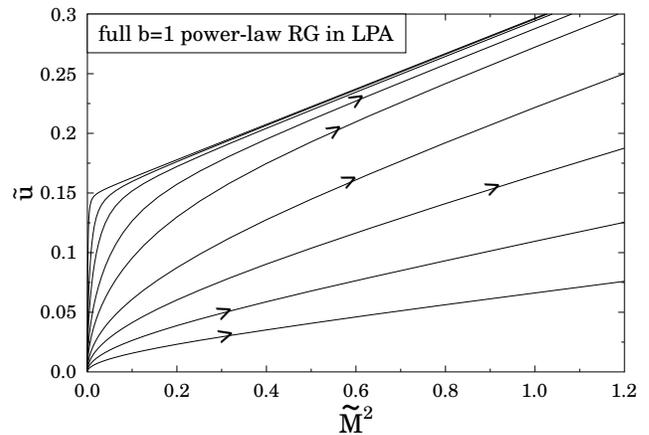,width=8.3 cm}
\caption{
\label{b1_phase}
Phase diagram of the MSG model for $\beta^2 =4\pi$. RG trajectories 
are obtained by the direct numerical solution of Eq.~\eq{b1} (solid lines) 
incorporating the effect of higher harmonics.
} 
\end{center}
\end{figure}

Indeed, the truncation of the expansion of the blocked potential in a series 
of base functions may become unreliable when the blocked action becomes 
almost degenerate, i.e. when $k^2+ V_k''$ approaches zero \cite{direct_rg}. 
This motivates a direct numerical solution of the RG Eq.~\eq{ea_v} for 
the blocked potential which avoids any assumption on the functional subspace 
where the solution is sought for and any truncated series expansion in some 
base functions. In this case, the appearance of SI is avoided and the critical 
ratio can be determined scheme-dependently. For example, Eq. \eq{ea_v} 
for the power-law type regulator with $b=1$ reads as
\beq
\label{b1}
(2 + k\partial_k) \tilde V_k(\varphi) =  
-\frac{1}{4\pi} \ln\left(1 + \tilde V''_k(\varphi) \right)
\eeq
of which direct integration results in $[\tu/\tilde M^2]_c = 0.148$ (see 
the solid lines in \fig{b1_phase}). 
%
%
Moreover, one obtains a better result 
$[\tu/\tilde M^2]_c = 0.159$ for the optimized regulator and also for the 
power-law type one with $b=2$.  Consequently, in case of the direct 
integration of RG equations (when the effect of higher harmonics are 
incorporated), (i) SI can be avoided in the RG flow, (ii) the critical 
ratio is found to be scheme-dependent, (iii) the RG schemes can be 
optimized using the known result \eq{exact_ratio} obtained for QED$_2$.
Let us note that the best result for the critical ratio (the maximum which 
can be achieved by any RG scheme in LPA) is obtained by regulators 
which have good convergence properties, e.g. in case of the optimized 
RG one has to choose $a=1$ and for the power-law RG the good choice is 
$b>1$. Those regulators which provide for RG equations poor convergence 
properties like the optimized regulator with $a\neq 1$ and the power-law 
type regulator with $b=1$ are not suitable to recover the best result for the 
critical ratio. For comparison, see \tab{tab1}.
%
%
\begin{table}[ht]
\small
\begin{center}
\begin{minipage}{8cm}
\begin{center}
\begin{center}
\begin{tabular}{|c|c|c|c|}
\hline
&&&\cr
opt. single-fr. & opt. multi-fr. & b=1 multi-fr. & b=2 multi-fr. \cr
&&&\cr
\hline
&&&\cr 0.062 & 0.159  &  0.148 & 0.159 \cr
\hline
\end{tabular}
\end{center}
\caption{
\label{tab1} 
Critical ratio obtained by various RG methods.}
\end{center}
\end{minipage}
\end{center}
\end{table}
%

\section{Functional RG beyond the LPA}
\label{sec_beyond_lpa}
Can the appearance of SI be avoided by the inclusion of the wave -unction 
renormalization? What is the IR value for the frequency (i.e. $\beta^2 = 1/z$)
in the broken symmetric phase of the MSG model if the wave-function 
renormalization is included? How does the wave-function renormalization 
affect bosonization? In order to clarify these issues, let us go beyond the 
LPA and solve the flow Eqs.~\eq{general_ea_u} and \eq{general_ea_z} 
obtained for the single-frequency MSG model where the wave-function 
renormalization $z$ is kept scale-dependent. Since the optimized regulator
does not support the derivative expansion beyond second order in this
section we focus on the power-law RG.

\begin{figure}[ht] 
\begin{center} 
\epsfig{file=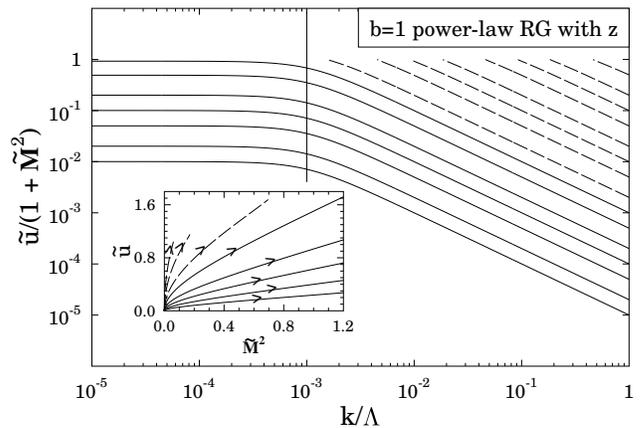,width=8.3 cm}
\caption{
\label{z_b1_phase}
RG trajectories are obtained by the integration of Eq. \eq{single_b1_exact}. 
Dashed lines correspond to RG trajectories where SI occurs, thus the critical 
ratio $r^{c}_{b=1}$ cannot be determined unambiguously. Vertical line shows the 
dimensionful mass scale which remains unchanged under RG transformations. 
} 
\end{center}
\end{figure}

Let us first consider the Callan-Symanzik scheme (i.e. the power-law RG with
$b=1$) where the flow Eqs.~\eq{general_ea_u} and \eq{general_ea_z} 
reduce to Eq. \eq{single_b1_exact}. As is seen in the inset of \fig{z_b1_phase}, 
SI appears in the RG flow in the broken symmetric phase, i.e. the RG equation 
becomes singular in the IR limit and the flow stops at some finite momentum 
scale (see the dashed lines in the inset of  \fig{z_b1_phase}).  
%
%

\begin{figure}[ht] 
\begin{center} 
\epsfig{file=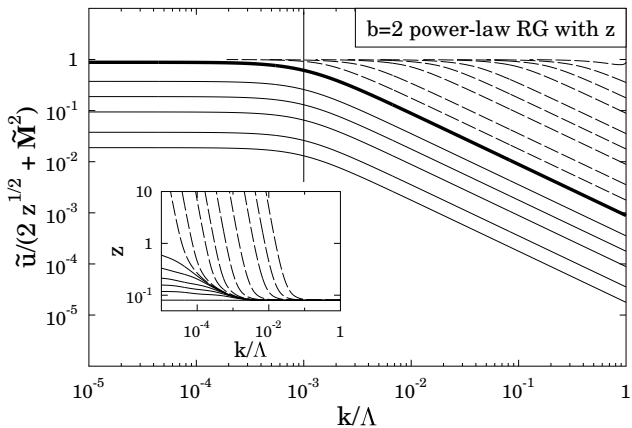,width=8.3 cm}
\caption{
\label{z_b2_phase}
RG trajectories are obtained by the numerical integration of Eqs. \eq{general_ea_u} 
and \eq{general_ea_z} for the power-law regulator with $b=2$. Dashed lines 
correspond to RG trajectories in the broken symmetric phase of the single-frequency 
MSG model. Vertical line shows the dimensionful mass scale which remains 
unchanged under RG transformations. The inset shows the scaling of the 
wave-function renormalization in the two phases. 
} 
\end{center}
\end{figure}

Since the convergence properties of the power-law RG are increased for 
$b>1$, Eqs. \eq{general_ea_u} and \eq{general_ea_z} are solved numerically 
for $b=2$ (see \fig{z_b2_phase}). Independently of the actual value of $b$,  the 
potential was found to become degenerate in the broken symmetric phase and 
the RG flow is determined by the degeneracy condition (similar results were 
obtained for the SG model in \cite{sg_prl})
\bea
c(b) z^{1-1/b} - {\tilde u} + \tilde {\bf M}^2 = 0
\eea
where $c(b) = b/(b-1)^{1-1/b}$. Therefore, in the IR limit the ratio  
\bea
\label{ratio_z}
r_{b}(k)  =  \frac{\tilde u}{c(b) z^{1-1/b} + \tilde {\bf M}^2}
\eea
tends to one (i.e. $r_{b}(k\to 0) = 1$) in the broken phase (see the dashed
lines  in \fig{z_b2_phase} for $b=2$). Therefore, the ratio becomes universal 
in the broken phase. In the symmetric phase it tends to a constant
IR value depending on the initial conditions. The critical ratio $r_{b}^{c}(k)$ 
which separates the phases of the single-frequency MSG model is represented
by the thick solid line in \fig{z_b2_phase}.
%
%
Similar results can be obtained for $b=1$ (see the dashed lines in \fig{z_b1_phase}),
but due to the poor convergence properties of the Callan-Symanzik scheme, SI
appears in the RG flow in the broken symmetric phase and the ratio cannot reach 
its universal value. Let us note that the RG flow always stops at a finite momentum 
scale in the broken phase independently of $b$ but a better convergence is 
obtained for $b>1$. This indicates that a more accurate calculation requires
the inclusion of higher harmonics as it was demonstrated in LPA. 

In general, the single-frequency approximation is "improved"  by the inclusion 
of the wave-function renormalization. For example, the critical exponent $\nu$ 
of the MSG model can be obtained in the framework of the power-law RG with
$b>1$ if $z(k)$ is kept scale-dependent. It is known \cite{dmrg_critical} that 
the MSG model belongs to the two-dimensional Ising universality class, thus 
the correlation length is a power-law function of the reduced temperature 
$\xi \sim t^{-\nu}$ with $\nu=1$. Indeed, if one defines the correlation length 
in the symmetric (disordered) phase by the constant IR values of the ratio,
$\xi \sim [1- r_{b}(k\to 0)]^{-1}$ and the reduced temperature is given by the 
initial UV ($k=\Lambda$) values,
$t = [r_{b}(\Lambda)^{-1} - r_{b}^{c}(\Lambda)^{-1}]/ r_{b}^{c}(\Lambda)^{-1}$
then $\nu = 1$ is obtained (see solid lines in \fig{z_b2_phase} for $b=2$). Let 
us note that in following Ref. \cite{sg_prl} the correlation length can be defined 
as $\xi \sim ({\bf M} - k_{c})^{-1}$ in the broken phase where $k_{c}$ 
represents the momentum scale at which the ratio \eq{ratio_z} becomes 
constant during the RG flow. If the reduced temperature is
$t = [r_{b}(\Lambda) - r_{b}^{c}(\Lambda)]/ r_{b}^{c}(\Lambda)$ then one 
obtains again the power-law behavior with $\nu =1$. Consequently, the RG 
equations derived for the single-frequency MSG model beyond LPA are
sufficient to indicate that the model undergoes a second order Ising-type
phase transition (while it is known that the SG model has an infinite order, 
KTB-type phase transition \cite{sg_prl}).

Let us consider the RG evolution of the wave-function renormalization
which is equivalent to the inverse frequency, i.e. $z(k) \equiv 1/\beta^2(k)$. 
In the symmetric phase $z(k)$ becomes a constant in the IR limit depending 
on the initial conditions (see the solid lines in the inset of \fig{z_b2_phase}).
In the broken symmetric phase, however $z(k)$ runs into infinity for $k \to 0$ 
(see dashed lines in the inset of \fig{z_b2_phase}), i.e. it has a universal 
behavior in the broken phase thus $\beta(k)$ tends to zero. Therefore, if 
one assumes that bosonization identifications between the parameters of 
the fermionic and the corresponding bosonic theory hold also for the 
blocked action then our result has a drawback on bosonization, namely 
it indicates the necessity to construct the fermionic counterpart of the 
MSG model for $\beta^2 \neq 4\pi$.

Finally, let us mention an open question related to the renormalization
of the MSG model. By the inclusion of the wave-function renormalization
it was possible to determine the critical exponent of the correlation 
function which demonstrates that the MSG model belongs
to the universality class of the two-dimensional Ising model i.e. the 
$O(N)$ symmetric scalar theory for $N=1$ and $d=2$. It is known that
higher order polynomial terms are needed in case of the $O(1)$ model
in order to obtain the critical behavior at the Wilson-Fisher fixed point in
a reliable manner. Thus, it might indicate that the critical behavior of the 
MSG model found at the IR limit is the consequence of a Wilson-Fisher 
type fixed point which could possibly appear in the MSG model if higher
polynomials of the field are incorporated in the RG flow. Therefore, it 
is an interesting open question to consider the phase structure of the
MSG model with the inclusion of higher order monomials of the field, 
although these terms are not generated by RG equations using the
ansatz \eq{msg}.

\section{Conclusions}
\label{sec_sum}
Known results on QED$_2$ have been used to optimize RG schemes for its 
bosonized version, the MSG model and to consider how the results obtained 
by RG equations depend on various approximations used. By the inclusion 
of the wave-function renormalization and the direct integration of RG equations 
derived for the MSG model, we went beyond the previously used approximations.

It was shown that the inclusion of higher harmonics and the direct integration 
of RG equations are both needed to avoid the appearance of singularity in 
the RG flow of the MSG model (for the optimized and power-law regulators) and 
to recover the critical ratio of QED$_2$. It is also demonstrated that the optimized 
RG predicts a reliable result for the single-frequency MSG model in LPA and it 
is known that the Wilson-Polchinski flow is not suitable to map out its phase 
structure \cite{scheme_sg}. Thus it shows that the two latter RG methods do not
produce the same critical behavior in LPA if the blocked potential becomes 
degenerate which is the case for the MSG model in its broken phase.

Moreover, as a result of optimization, the best result for the critical ratio of QED$_2$
(the maximum which can be reached in LPA) is obtained by those regulators 
which have good convergence properties, such as the optimized RG with $a=1$
and the power-law RG with $b>1$. Regulators with poor convergence properties
(e.g. optimized RG with $a>1$ and power-law RG with $b=1$) are found to be
unable to recover the best result.

If one assumes that the identifications between the parameters of the fermionic 
and the corresponding bosonic theory holds also for the blocked action then 
results on the MSG model indicate that the renormalization of QED$_2$ 
possibly requires interaction terms which correspond to higher frequency 
modes of its bosonized version. The renormalization of the wave-function in the 
RG flow of the MSG model has also a drawback on bosonization. Since in this 
case $z=1/\beta^2$ is scale-dependent, it is a necessity to construct the fermionic 
counterpart of the MSG model for $\beta^2\neq 4\pi$. Indeed, if one assumes a 
quartic self-interaction among the massive Dirac fermions of QED$_2$ by adding 
a Thirring-type term to the Lagrangian \eq{qed_2} then one arrives at the massive 
Schwinger-Thirring model \eq{mst} which was proposed as the corresponding 
fermionic theory of the MSG model for $\beta^2 \neq 4\pi$.

Finally, let us note that the scenario discussed in this work can possibly be 
extended for many-body condensed matter systems \cite{bose_higher_dim}. 
Another open question related to the present work is the direct comparison 
between flows of fermionic and bosonic models. For example, one can try 
to compare the functional RG study of the SG scalar theory and the fermionic
Thirring model which can possibly be achieved by the extension of the RG 
analysis of the three-dimensional Thirring model \cite{3d_thirring} to the 
two-dimensional one. Furthermore, bosonized versions of multiflavor 
QED$_2$ and mutlicolor QCD$_2$ are also SG-type models, hence they 
represent further examples where the drawback of RG results can be studied 
on bosonization (including wave-function renormalization as the frequency 
becomes scale-dependent).

\section*{Acknowledgement}
This research was supported by the T\'AMOP 4.2.1./B-09/1/KONV-2010-0007 project.

\end{document}